# Metal surface structuring with spatiotemporally focused femtosecond laser pulses

Yuanxin Tan[1,2], Wei Chu[1], Jintian Lin[1], Zhiwei Fang[1,2,3], Yang Liao[1], Ya Cheng[1,4,5]

[1] State Key Laboratory of High Field Laser Physics, Shanghai Institute of Optics and Fine Mechanics, Chinese Academy of Sciences, Shanghai 201800, China.
[2] University of Chinese Academy of Sciences, Beijing 100049, China.
[3] School of Physical Science and Technology, ShanghaiTech University, Shanghai 200031, China
[4] State Key Laboratory of Precision Spectroscopy, East China Normal University, Shanghai 200062, China.
[5] Collaborative Innovation Center of Extreme Optics, Shanxi University, Taiyuan, Shanxi 030006, China.

Correspondence and requests for materials should be addressed to W. C. (email: chuwei0818@qq.com) or Y.C. (email: ya.cheng@siom.ac.cn).

**Abstract**

Femtosecond laser micromachining provides high precision and less thermal diffusion in surface structuring as a result of the ultrashort temporal duration and ultrahigh peak intensity of the femtosecond laser pulses. To increase the throughput of surface patterning, the focal spot size can be expanded with loose focusing, which, however, could lead to nonlinear self-focusing of the pulses when the pulses propagate in air. We solve the problem by use of spatiotemporally focused femtosecond laser pulses for ablation of metal surfaces, which gives rise to improved surface quality as compared with that obtained with the conventional focusing scheme.

Laser surface structuring is one of the major applications of laser materials processing, providing advantages of flexibility in patterning on surfaces of various materials with arbitrary geometries and topographies and environment friendliness [1-3]. Recently, the rapid progress in femtosecond laser materials processing makes femtosecond laser surface structuring an attractive means to achieve high quality material removal from the surface of almost any kind of materials [4-9]. The reason is that due to the ultrashort pulse durations, the interaction between the femtosecond laser and the materials can be regarded as a "cold" process for the reason that before the thermal diffusion, which typically takes hundreds of femtoseconds or a few picoseconds, substantially occurs, the laser pulses have already terminated [10]. This opens the avenue to the high quality surface structuring including nanostructuring [11, 12], patterning [13, 14], dicing [15], to name a few.

From an industrial point of view, the pointwise fashion of laser direct writing inevitably leads to low fabrication throughputs. The problem can be partially solved by increasing the number of focal spots (e.g., an array of focal spots) or the repetition rate of the laser pulses. The former is a powerful approach to promote the fabrication efficiency provided that the patterns to be fabricated are periodic structures; otherwise this approach lacks the flexibility in terms of producing patterns of arbitrary geometries without a periodicity. The latter maintains the geometric flexibility of the direct writing and can efficiently promote the throughput by greatly raising the repetition rate of the femtosecond laser. Nevertheless, even in the high-repetition-rate writing case, the fabrication throughput can be further increased by enlarging the focal spot size, provided that the feature size of structures to be produced on the surface is not as small as the scale of wavelength.

Traditionally, the focal spot size can be controlled by tuning the numerical aperture (NA) of the focal lens which is inversely proportional to the diameter of focal spot. However, in the femtosecond laser processing, this strategy can be problematic as loosely focused femtosecond laser pulses can easily self-focus to form a filament elongated along the propagation direction before the arrival of pulses at the geometrical focus, which leads to distortion of the focal spot at the focus. From the materials processing point of view, it is often needed to eliminate the self-focusing synergetically contributed by the loose focusing and high peak intensity of the laser pulses. Here, we show that this can be achieved using spatiotemporal focusing (STF) of the femtosecond laser pulses [16-19]. We further demonstrate that when performing the metal surface structuring in air, the STF scheme will produce high quality surfaces showing less influence from thermal effects than that of the metal surfaces processed with the conventionally focused laser pulses.

## Results and Discussion

Figure 1(a) and (b) illustrates the surface ablation systems with conventional scheme and STF scheme, respectively. The major difference in the two setups is that in the STF scheme, a pair of gratings are used to produce a spatial chirp of the incident pulses at the back aperture of the focal lens, as indicated as grating 1 and grating 2 in Fig. 1(b). The technical details can be found in **Methods**. The striking difference in focusing the femtosecond laser pulses in air with the different focal systems is emphasized in Fig. 2. Here, when the femtosecond pulses of a pulse energy of 80 µJ were focused using the STF scheme with a focal lens of focal distance f = 200 mm, we could not observe any visible signature of optical breakdown in air as shown in Fig. 2(a). This indicates that the peak intensity at the focus of the focal lens is insufficient to generate photoionization because of the low NA value (NA=0.01). Assuming a linear propagation under the focusing condition, the diameter of the focal spot should be ~50 μm at the focus, which is significantly larger than the wavelength of the laser pulses. However, when switching to the conventional focusing scheme by maintaining the same NA value of the focal system, we observed a long filament of a length of ~4 mm from the side of optical axis of the focal lens, as indicated by the spatial profile of the fluorescence from the photoionized nitrogen molecules in air in the inset of Fig. 2(b). The filament is essentially a plasma channel formed in the laser field by the combined effort of the nonlinear self-focusing and the plasma-induced defocusing. The result reveals that to loosely focus a femtosecond laser beam for the purpose of increasing the focal spot size, the self-focusing during the propagation of the beam in air will cause an optical breakdown prior to the arrival of the pulses onto the target. The optical breakdown will spoil the spatial and temporal properties of the laser pulses and can often be sensitive to the fluctuation of the initial conditions of the laser beam (e.g., pulse duration, pulse energy, pointing stability, spatial profile, etc.). Thus, the optical breakdown in air before the geometric focus always jeopardizes the stability in the materials processing in such circumstances. However, the self-focusing can be greatly suppressed with the STF scheme, as evidenced in Fig. 2(a) where no any signature of optical breakdown is seen. Next, we will examine this concept by performing the ablation on the surfaces of several metals with both focusing schemes and compare the qualities of ablated surfaces.

Figure 3 compares the ablation results on the surface of copper, with the surfaces produced with the conventional focusing scheme presented in Fig. 3(a-c) and that produced with the STF scheme presented in Fig. 3(d-f). The pulse energy increases from the upper to the lower row, of which the values are directly indicated in the caption of Fig. 3. In all the fabrication process, the writing speed was fixed at 2.5 mm/s. It can be observed that with an increasing pulse energy from 30 μJ to 80 μJ, the surface roughness becomes worse for surfaces presented in Fig. 3(a-c), whilst the surfaces in Fig. 3(d-f) do not significantly change their morphology and show much less influence from the thermal effects. We would like to point out that when we increase the pulse energy from 30 μJ to 80 μJ, the ablation becomes much more efficient, and we have observed that the depth of the ablated trench increases from ~18 μm to ~38 μm. The ablation depths obtained with STF and conventional focal schemes are very similar without significant quantitative difference. The high ablation rate is, in many circumstances, highly demanded by the industrial applications.

The comparison of the ablation between conventional focusing and STF scheme on the surface of aluminum was also demonstrated, with the surfaces produced with the conventional focusing scheme presented in Fig. 4(a-c) and that produced with the STF scheme presented in Fig. 4(d-f). Likewise, when the pulse energy increases from 20 μJ to 60 μJ, the ablation appears to be much less influenced by the thermal effects with the STF scheme than that with the conventional focusing scheme. As the pulse energy increases from 20 μJ to 60 μJ, the measured depth of the ablated trench increases from ~17 μm to ~40 μm, respectively. Again, the ablation depths obtained with STF and conventional focal schemes are similar to each other.

At last, the results of ablation on the surface of stainless steel are compared in Fig. 5. The depth of the ablated trench increases from ~15 μm to ~40 μm with the increase of the processing pulse energy from 20 μJ to 60 μJ for both focusing schemes. The parameters for producing the results in Fig. 5(a-f) are given in the caption of the figure. Obviously, the results obtained with the STF scheme are better than that obtained with the conventional focusing in terms of surface quality. On the other hand, the morphologies on the surface of stainless steel obtained at the high pulse energies are worse than that on the surfaces of copper and aluminum obtained under the same ablation conditions.

The results given above show that the STF scheme can help improve the quality for metal surface structuring particularly under high peak laser intensities, a condition frequently required for high-throughput surface structuring. One of the physical reasons behind this is shown in Fig. 2. One can realize that when ablate the surface with a conventional focusing scheme, plasma will be formed in air before the arrival of the pulses onto the surfaces of the targets. In such a case, the interaction between the metal surfaces with the laser pulses is complicated, as the plasma generated in air will participate in the ablation process. Several scenarios could be expected as follows. First, the optical breakdown in air could lead to extra instability in the ablation due to the sensitivity of highly nonlinear photoionization on the fluctuation of laser parameters such as pulse energy, pulse duration and pointing directions, etc [20]. Second, the plasma generated before the target will change the properties of the focal spot on the target, such as the peak intensity, focal spot size, and pulse duration, etc [21]. Third, the plasma generated in air will also influence the plasma dynamics of the ablation process, as the plasma generated at the surfaces may be unable to efficiently eject into the ambient environment due to the existence of pre-formed plasma in air. For this reason, re-deposition of ablated materials on the surfaces can be clearly observed on the surfaces ablated with the conventional focusing scheme. All of these problems can be greatly mitigated in the ablation with the STF scheme due to the suppressed plasma generation in air, as the spatiotemporally focused pulses undergo the least distortion on the targets for producing the high-quality surface ablation.

Another interesting feature is that the surface morphologies in Figs. 3, 4 and 5 are different for different metal materials even under the same ablation conditions. The physics behind this observation is rather unclear and provides a nice topic of research to be systematically investigated in the future. Nevertheless, generally speaking, the physics has to be related to the different material properties of the metals as has been discussed in Ref. 22. The three metals investigated in this work have different thermal conductivities and electron-phonon coupling strengths as shown in Tab. 1 [22, 23]. For metals such as copper and aluminum which have smaller electron-phonon coupling strengths, the absorbed laser energy will be more likely transferred into the bulk but not localized at the surfaces, as the hot electrons can diffuse into the bulk more easily owing to the high thermal conductivities of these metals. Such process reduces the thermal effects on the surfaces, giving rise to the formation of cleaner surface topographies. Switching to stainless steel which is of a smaller thermal conductivity and a larger electron-phonon coupling strength, the absorbed laser energy is more likely localized near the surface, leading to more pronounced heat affected zones.

## Conclusion

To conclude, we have investigated the surface ablation on different metals with both the STF scheme and the conventional focusing scheme. Our results show different characteristics in the ablated surfaces with the two schemes. The STF allows for producing high-quality surfaces without much signature of thermal influences even at the loose focusing condition which gives rise to large focal spot sizes and high peak intensities for promoting the manufacturing throughputs. Thus, our result is vital for many industrial applications. Further effort will be made on systematic investigations on the physics behind the different surface morphologies obtained with the two schemes and on the femtosecond laser patterning surfaces of dielectric materials (e.g., glass, crystals, polymers, etc.) with the STF scheme.

## Methods

**Experimental details**

In this study, we used an amplified Ti:sapphire laser system (Libra-HE, Coherent Inc.) that generates femtosecond laser pulses with a central wavelength of 800 nm and at a pulse repetition rate of 1 kHZ. In the conventional focusing scheme, the 40-fs transform-limited pulses from the compressor were firstly passed through a telescope system consisting of a convex lens (f = 400 mm) and a concave lens (f = −200 mm) to reduce the beam width to ~4 mm ($1/e^2$), subsequently reflected by golden mirrors and dichromatic mirror, and then focused onto the metal surfaces by a focusing lens (f = 200 mm). A charged coupled device (CCD) was installed above the objective lens and the dichromatic mirror to monitor the surface processing in real time, as shown in Fig.1 (a). For our STF configuration, the uncompressed femtosecond laser pulses from the amplifier passed through the single-pass grating compressor consisting of two σ = 1500 grooves/mm gratings, blazed for the incident angle of 53°. The distance between the two gratings was adjusted to compensate for the temporal dispersion of pulses. After being dispersed by the grating pair, the spatially dispersed laser pulses were then focused onto the surface of the metals, as shown in Fig 1 (b).

The copper, aluminium and stainless steel 304 (HF-Kejing materials technology CO., LTD.) used in the investigation were diced into small coupons of 1 cm$^2$ in size and 1 mm in thickness. The purities of the copper and aluminium are 99%. All the surfaces were polished and the samples were cleaned in an ultrasonic bath with acetone before being processed with the femtosecond laser surface ablation.

.

## Acknowledgments

This work is supported by the National Basic Research Program of China (Grant No. 2014CB921303), National Natural Science Foundation of China (Grant Nos. 61327902, 11674340, 61505231, 61575211, 61675220, 61590934, 11604351 and 11404357) and Strategic Priority Research Program of the Chinese Academy of Sciences, Grant No. XDB16030300.

## Author contributions

Y. C. and W. C. planned and designed the experiments. Y. T. and W. C. performed the surface ablation experiment. Y. T., J. L, Z. W. and W. C. observed the surface structure by scanning electron microscopy. Y. C., W. C. and Y. T. wrote the paper. All authors participated in the discussion of the results.

## Additional information

**Competing financial interests:** The authors declare no competing financial interests.

Tab 1. Thermal conductivities and electron-phonon coupling constants for copper, aluminum, stainless steel. The values of thermal conductivities are given for room temperature (i.e., 300 K).

| Metal | Thermal conductivities (W/m/K) | Electron-photon coupling strengths($10^{17}$ W/m$^3$/K) |
|---|---|---|
| Copper | 401 | 0.48±0.07 |
| Aluminum | 177 | 2.45±0.14 |
| Stainless steel | 14.9 | 30 |

**Figure Captions**

Figure 1. Schematic illustrations of femtosecond laser surface structuring. (a) Conventional focusing scheme. (b) STF scheme. GM: golden mirror. DM: dichromatic mirror.

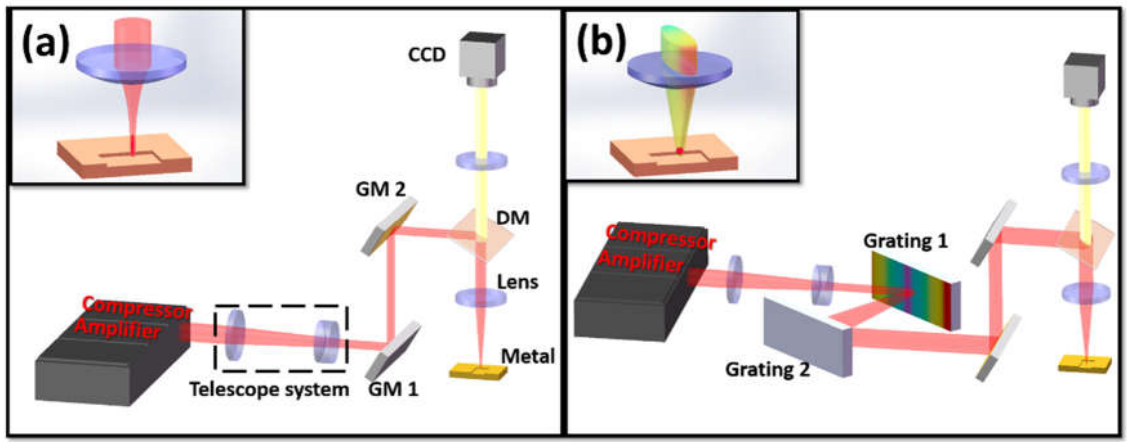

Figure 2. The digital-camera-captured pictures (integrated time: 30s) of the optical breakdown in air for the different focusing schemes: (a) STF scheme (where actually no breakdown occurs); (b) Inset in (b): The spatial distribution of the fluorescence intensity along the propagation direction.

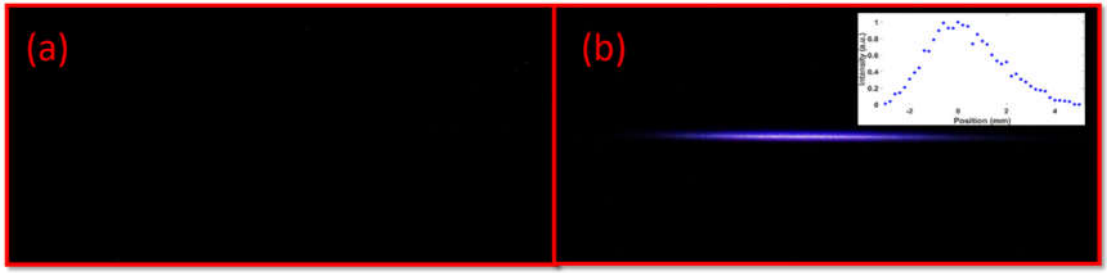

Fig. 3. The SEM images of ablated surfaces of copper with (a-c) conventional focusing scheme, and (d-f) STF scheme. Pulse energy: (a, d) 30 µJ; (b, e) 50 µJ; (c, f) 80 µJ.

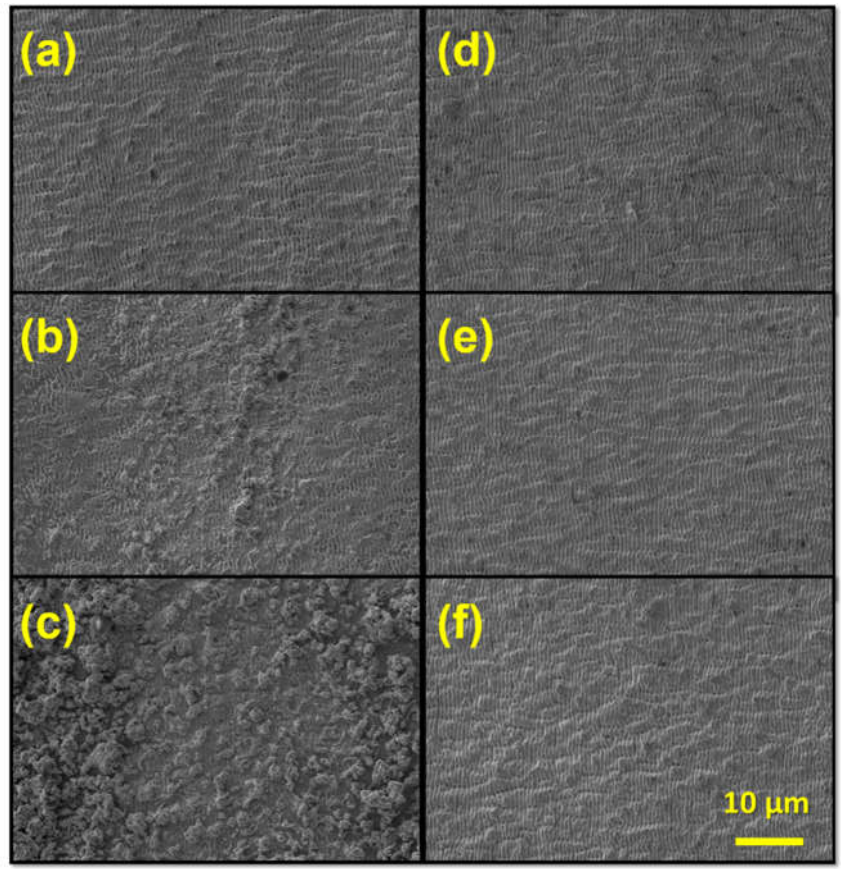

Fig. 4. The SEM images of ablated surfaces of aluminum with (a-c) conventional focusing scheme, and (d-f) STF scheme. Pulse energy: (a, d) 30 μJ; (b, e) 50 μJ; (c, f) 80 μJ.

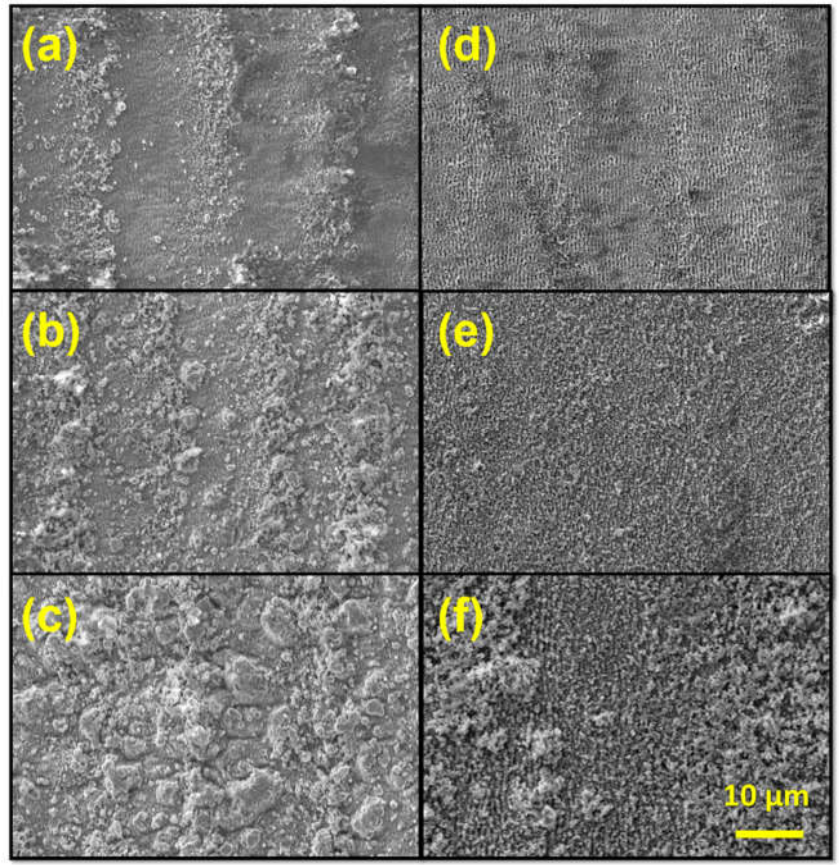

Fig. 5. The SEM images of ablated surfaces of stainless steel with (a-c) conventional focusing scheme, and (d-f) STF scheme. Pulse energy: (a, d) 30 μJ; (b, e) 50 μJ; (c, f) 80 μJ.

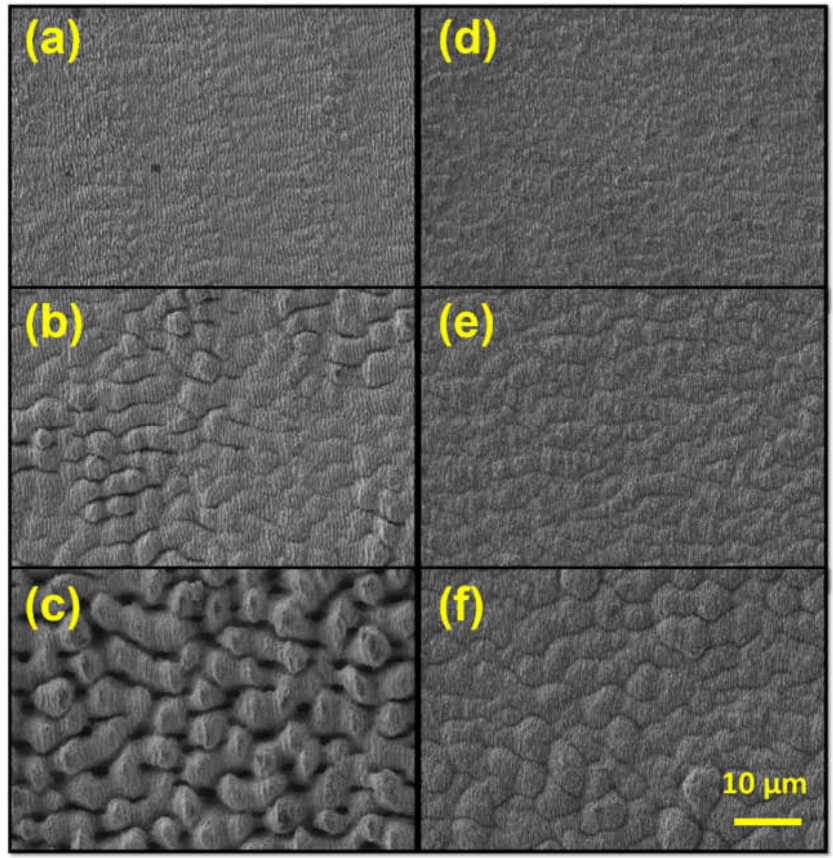